\documentclass[journal,article,submit,moreauthors,pdflatex,12pt,a4paper]{mdpi} 
\lastpage{x}
\doinum{10.3390/------}
\pubvolume{xx}
\pubyear{2012}
\history{Received: xx / Accepted: xx / Published: xx}

\usepackage{amsmath}
\usepackage{amssymb}

\Title{Casimir Friction Between Dense Polarizable Media
}

\Author{Johan S. H{\o}ye $^{1}$ and Iver Brevik $^{2,\star}$}

\address{%
$^{1}$ Department of Physics, Norwegian University of Science and Technology, N-7491 Trondheim, Norway\\
$^{2}$ Department of Energy and Process Engineering, Norwegian University of Science and Technology, N-7491 Trondheim, Norway
}

\corres{E-Mail iver.h.brevik@ntnu.no, Tel. +47 7359 3555,  Fax +47 7359 3491.}

\abstract{The present paper - a continuation of our recent series of papers on Casimir friction for a pair of particles at low relative particle velocity - extends the analysis so as to include dense media. The situation becomes in this case more complex due to induced dipolar correlations, both  within planes, and between planes. We show that the structure of the problem can be simplified by regarding the two half-planes as a generalized version of a pair of particles. It turns out that  macroscopic parameters such as permittivity suffice to describe the friction also in the finite density case. The expression for the friction force per unit surface area becomes mathematically well-defined and finite at finite temperature. We give numerical estimates, and compare them with those obtained earlier by Pendry (1997) and by Volokitin and Persson (2007). We also show in an appendix how the statistical methods that we are using, correspond to the field theoretical methods more commonly in use.}


\begin{document}

\section{Introduction}

\label{sec1}

The typical situation envisaged in connection with Casimir friction is the one where  two parallel semi-infinite dielectric nonmagnetic plates at micron or semi-micron separation are moving longitudinally  with respect to each other, one plate being at rest, the other having a nonrelativistic velocity $\bf v$. Usually the plates are taken to have the same composition, their permittivity  $\varepsilon(\omega)$  being frequency dependent.

Most previous works on Casimir friction are formulated within the framework of macroscopic electrodynamics. Some references in this direction are \cite{teodorovich78,pendry97,pendry98,pendry10,volokitin99,volokitin03,volokitin07,volokitin08,volokitin11,dedkov08,dedkov10,dedkov11,dedkov12,philbin09}. In particular, the application of the theory to graphene materials is a very promising avenue of approach; cf., for instance, Ref.~\cite{volokitin11}. In the present paper we focus on the following themes:

\bigskip

$\bullet$ We make use of {\it statistical mechanical methods} for harmonic oscillators,  moving
with respect to each other with constant velocity $\bf v$, at a finite temperature $T$.  We claim that such a strategy, formally perhaps simpler than field-theoretical methods, is actually quite powerful. We have used this method repeatedly in previous recent  investigations \cite{hoye10,hoye11A,hoye11B,hoye12A,hoye12B}; cf. also the earlier  papers \cite{hoye92,brevik88} in which the foundations of the method were spelled out. The essence of the method is to generalize the statistical mechanical Kubo formalism to time-dependent cases.

\bigskip
$\bullet$ These methods are then used to generalize the theory to the case of {\it dense} media. This is a nontrivial task,  as the additivity property   holding for dilute media is no longer valid. This topic is dealt with from Sect. 4 onwards. One will have to deal with a more complicated form of the Green function. The atomic polarizabilities appearing in the theory of dilute media have to be replaced by by functions based upon the frequency dependent permittivity. A noteworthy property is, however, that the permittivity, i.e. a macroscopic quantity, suffices to express the Casimir friction even in the case of finite densities.

\bigskip
$\bullet$ It turns out that  the friction force becomes finite at finite temperature,  although usually small. As a numerical example, treated in Sect. 5, we find for the case of a gold metal  that the force per unit surface area for equal plates at room temperature at small separation (10 nm) and moderate relative velocity (100 m/s) becomes of order  $10^{-11}$ Pa. This, of course, cannot be measured. However, by changing input parameters this will change rapidly so that $F$ can become large. The situation is very sensitive with respect to input parameters. We make numerical comparisons with earlier works, notable Pendry (1997), and Volokitin and Persson (2007).

\bigskip
$\bullet$ As it is of interest to trace out the connection with the more standard field theoretical methods, we focus on this subject in Appendix B.  Appendix A shows or indicates the formal background for the correlation functions used.

\bigskip

 We mention that the microscopic approach has been followed by other investigators also, especially by Barton \cite{barton10A,barton10B,barton11}. The equivalence between our approaches, actually a rather  a nontrivial correspondence,  has been shown by us explicitly  \cite{hoye11B}.

For reasons of readability we begin in the next section by summarizing essential points of the theory for  dilute media \cite{hoye12A,hoye12B}. Then, after giving an account of Fourier methods in Sect. 3 we embark, as mentioned, on the general case of finite particle density in Sect. 4.


\section{Dilute Media}

 \label{sec2}

 For a pair of  polarizable particles the electrostatic dipole-dipole pair interaction perturbs the Hamiltonian by an amount
\begin{equation}
-AF(t)=\psi_{ij}s_{1i}s_{2j}, \label{1}
\end{equation}
where the summation convention for repeated indices $i$ and $j$ is implied. The $s_{1i}$ and $s_{2j}$ are  components of the fluctuation dipole moments of the two particles $(i,j=1,2,3)$. With  electrostatic dipole-dipole interaction we can write
\begin{equation}
\psi_{ij}=-\frac{\partial^2}{\partial  x_i\partial x_j}\psi, \quad \psi=\frac{1}{r}, \label{2}
\end{equation}
(i.e. $\psi_{ij}=-(3x_ix_j/r^5-\delta_{ij}/r^3)). $ Here ${\bf r}={\bf r}(t)$ with components $x_i=x_i(t)$ is the separation between the particles. The time dependence in Eq.~(\ref{1}) is due to the varation of $\bf r$ with time $t$, and the interaction will vary as

\begin{equation}
-AF(t)=\left[ \psi_{ij}({\bf r}_0)+\left(\frac{\partial}{\partial x_l}\psi_{ij}({\bf r}_0)\right)v_l t+...\right]s_{1i}s_{2j}, \label{3}
\end{equation}
where $v_l$ are the components of the relative velocity $\bf v$. The components of the force $\bf B$ between the oscillators are
\begin{equation}
B_l= -T_{lij}s_{1i}s_{2j}, \quad T_{lij}=   \frac{\partial}{\partial x_l}\psi_{ij}. \label{4}
\end{equation}
The friction force is due to the second term of the right hand side of Eq.~(\ref{3}), and for dilute media the first term can be neglected. However, for the more general situation to be considered below, this will no longer be the case since correlations will be induced.

For the time dependent part of Eq.~(\ref{3}) we may write $-AF(t)\rightarrow -A_lF_l(t)$ where $A_l=B_l$ and $F_l(t)=v_lt.$ According to Kubo \cite{kubo59,brevik88,hoye92} the perturbing term leads to a response in the thermal average of $B_l$ given by
\begin{equation}
\Delta \langle B_l(t)\rangle=\int_{-\infty}^\infty \phi_{BAlq}(t-t')F_q(t')dt', \label{5}
\end{equation}
where the response function is ($t>0 $)
\begin{equation}
\phi_{BAlq}(t)=\frac{1}{i\hbar}{\rm{Tr}}\,\{ \rho[A_q, B_l(t)]\}. \label{6}
\end{equation}
Here $\rho$ is the density matrix and $B_l(t)$ is the Heisenberg operator $B_l(t)=e^{itH/\hbar}B_l\,e^{-itH/\hbar}$ where $B_l$ like $A_q$ are time independent operators. With Eqs. (\ref{3}) and (\ref{4}) the  expression (\ref{6})  can be rewritten as
\begin{equation}
\phi_{BAlq}(t)=G_{lqijnm}\phi_{ijnm}(t), \label{7}
\end{equation}
where
\begin{equation}
G_{lqijnm}=  T_{lij}T_{qnm},      \label{8}
\end{equation}
\begin{equation}
\phi_{ijnm}(t)={\rm Tr} \{\rho C_{ijnm}(t)\}, \label{9}
\end{equation}
\begin{equation}
C_{ijnm}(t)=\frac{1}{i\hbar}\left[ s_{1i}s_{2j}, s_{1n}(t) s_{2m}(t)\right] \label{10}
\end{equation}
(the $i$ in the denominator is the imaginary unit).

Here as in Ref.~\cite{hoye12A} it is convenient to use imaginary time $\lambda$  and consider the correlation function
 \begin{equation}
g_{ijnm}(\lambda)={\rm Tr}[\rho s_{1n}(t)s_{2m}(t)s_{1i}s_{2j}], \label{11}
\end{equation}
with $\lambda=it/\hbar$.

The key problem when dealing with media with general permittivity will be the evaluation of this function to obtain the friction force. For dilute media, however, the two oscillators (assumed isotropic) are independent, so we have
\begin{equation}
g_{ijnm}(\lambda)=  g_{1in}(\lambda)g_{2jm}(\lambda),  \label{12}
\end{equation}
\begin{equation}
g_{apq}(\lambda)=\langle s_{aq}(t)s_{ap}\rangle =g_a(\lambda)\delta_{qp}, \quad (a=1,2), \label{13}
\end{equation}
where the angular brackets denote thermal averages $( \langle ..\rangle={\rm Tr} [\rho ..])$.

The $\phi$ is related to the $g$ via
\begin{equation}
 \phi_{ijnm}(t)=\frac{1}{i\hbar}\left[ g_{ijnm}(\beta+\lambda)-g_{ijnm}(\lambda)\right], \label{14}
 \end{equation}
 and
 \begin{equation}
 \tilde{\phi}(\omega)=\tilde{g}(K), \label{15}
 \end{equation}
 where the Fourier transforms are
 \begin{equation}
  \tilde{\phi}(\omega)=\int_0^\infty \phi(t)e^{-i\omega t}dt, \label{16}
  \end{equation}
\begin{equation}
\tilde{g}(K)=\int_0^\beta g(\lambda)e^{iK\lambda} d\lambda, \label{17}
\end{equation}
 with $K$ the imaginary frequency,
 \begin{equation}
 K=i\hbar \omega. \label{18}
 \end{equation}
 Here  $\beta=1/k_B T$, where $T$ is the temperature and $k_B$  Boltzmann's constant.

 With Eqs.~(\ref{12}) and (\ref{13}) we have
 \begin{equation}
 g_{ijnm}(\lambda)=     g(\lambda)\delta_{in}\delta_{jm},  \label{19}
 \end{equation}
 \begin{equation}
 g(\lambda)=      g_1(\lambda)g_2(\lambda), \label{20}
 \end{equation}
 by which the $\tilde{g}(K)$ can be written as a convolution
 \begin{equation}
{\tilde g}(K)=\frac{1}{\beta}\sum_{K_0}{\tilde g}_1(K_0){\tilde g}_2(K-K_0), \label{21}
\end{equation}
$K_0=2\pi n/\beta$ ($n$ is integer) being the Matsubara frequencies.
The $\tilde{g}_a(K)$ can be identified with the frequency dependent polarizability $\alpha_{aK}$ of oscillator $a$ (=1,2), which for a simple harmonic oscillator is
 \begin{equation}
{\tilde g}_a(K)=\alpha_{aK}=\frac{\alpha_a(\hbar \omega_a)^2}{K^2+(\hbar \omega_a)^2}, \label{22}
\end{equation}
where  $\alpha_a$ is the zero-frequency polarizability.

With (\ref{15}) and (\ref{19}) the Fourier transform of the expression (\ref{9}) will be
\begin{equation}
{\tilde \phi}_{ijnm}(\omega)={\tilde \phi}(\omega)\delta_{in}\delta_{jm}, \label{23}
\end{equation}
 where we recall that $\tilde{\phi}(\omega)=\tilde{g}(K)$.

 Further following Ref.~\cite{hoye12A} the friction force is given by
 \[
F_{fl}=-G_{lq}v_q\int_0^\infty \phi(u)udu = -iG_{lq}v_q\frac{\partial {\tilde \phi}(\omega)}{\partial \omega}\Big|_{\omega=0} \]
\begin{equation}
 = -G_{lq}v_q H\frac{\pi \beta}{2}\delta(\omega_1-\omega_2), \label{24}
 \end{equation}
 where
 \begin{equation}
 G_{lq}=G_{lqiijj}=T_{lij}T_ {qij}, \label{25}
 \end{equation}
 \begin{equation}
 H=\left( \frac{m}{2\sinh (\frac{1}{2}\beta m)}\right)^2 \alpha_1\alpha_2, \label{26}
 \end{equation}
 with $m=\hbar \omega$, where $\omega_1-\omega_2=\omega$.

 The treatment above can be extended to a more general polarizability
 \begin{equation}
 \alpha_a(K)=\tilde{g}_a(K)=f(K^2), \label{27}
 \end{equation}
 where it can be shown  that the function $f(K^2)$ satisfies the relation \cite{hoye82}
 \begin{equation}
 f(K^2)=\int \frac{\alpha_{Ia}(m^2)m^2}{K^2+m^2}d(m^2), \label{28}
 \end{equation}
 with
 \begin{equation}
 \alpha_{Ia}(m^2)m^2=-\frac{1}{\pi}{\Im}[f(-m^2+i\gamma)], \quad (m=\hbar \omega=-iK,~\gamma \rightarrow 0^+). \label{29}
 \end{equation}
 With this one finds
 \begin{equation}
 F_{fl}=-G_{lq}v_qH_0, \label{30}
 \end{equation}
 \begin{equation}
 H_0=\frac{\pi \beta\hbar^2}{2}\int \frac{m^4\alpha_{I1}(m^2)\alpha_{I2}(m^2)}{\sinh^2(\frac{1}{2}\beta m)} d\omega. \label{31}
 \end{equation}
This is obtained by replacing $\alpha_a$ with $\alpha_{Ia}(m_a^2)\,d(m_a^2)$ ($a=1,2$) in expression (\ref{26}) which is then inserted in Eq.~(\ref{24}) and then integrated with the $\delta$-function included.

 Finally by integrating $G_{lq}$ over space one obtains for dilute media the friction force $F_h$ between a particle and a half-plane, and the friction force $F$ (per unit area) between two half-planes that move parallel to each other
 \begin{equation}
 F_h=-G_h vH_0, \quad {\rm and} \quad F=-GvH_0. \label{32}
 \end{equation}
 Here $v$ is the relative velocity in the $x$ direction, and one finds \cite{hoye12A}
 \begin{equation}
 G_h=\rho_1\int_{z>z_0}G_{11}dxdydz=\frac{3\pi \rho_1}{2z_0^5}, \label{33}
 \end{equation}
 \begin{equation}
 G=\rho_2\int_d^\infty G_hdz=\frac{3\pi}{8d^4}\rho_1\rho_2. \label{34}
 \end{equation}
 Here $\rho_1$ and $\rho_2$ are the particle densities in the half-planes, $z_0$ is the separation between the particle and one half-plane, and $d$ is the separation between the half-planes.


\section{Use of Fourier Methods}

 For higher densities of polarizable particles the above results will be modified due to  induced dipolar correlations within planes, and between planes. This affects the evaluations of $G_h$ and $G$ which become more complex and demanding. Further, the expression (\ref{31}) for $H_0$ will be modified where we will find that the imaginary parts of the polarizabilities will be replaced by functions based upon the frequency dependent permittivity only. Thus the friction will depend solely on this macroscopic property, and be independent of the explicit relation between the permittivity and the polarizability. On physical grounds we find this reasonable.

 To facilitate the analysis we find it convenient to evaluate the integral (\ref{33}) by use of a Fourier transform in the $x$- and $y$-directions. Then the quantities in (\ref{2}) - (\ref{4}) should be transformed. The three-dimensional Fourier transform of the Coulomb potential $\psi=1/r$ is
 \begin{equation}
 \tilde{\psi}(k)=\frac{4\pi}{k^2}, \quad k^2=k_ik_i, \label{35}
 \end{equation}
 where $k_i~(i=1,2,3$ or $x,y,z$) is the Fourier variable. This can be transformed back with respect to $z$ to obtain
 \begin{equation}
 \hat{\psi}(z,k_\perp)=\frac{2\pi e^{-ik_zz}}{k_\perp}=\frac{2\pi e^{-q|z|}}{k_\perp}, \label{36}
 \end{equation}
 where $q=k_\perp, k_\perp^2=k_x^2+k_y^2$, and $ik_z=\pm q$ for $z >0$ or $z<0$. The variable $k_z$ may seem unnecessary here, but it is kept for convenience in order to  obtain simple and compact expressions for the transforms of derivatives. Otherwise one would need separate expressions for the transforms in the $x$ and $y$ directions and in the $z$ direction.
 
 Thus with Eqs. (\ref{2}) and (\ref{4}) $(\partial /\partial x_j \rightarrow -ik_j)$
 \begin{equation}
 \hat{\psi}_{ij}=\hat{\psi}_{ij}(z, {\bf k}_\perp)=-k_ik_j\hat{\psi}, \nonumber
 \end{equation}
 \begin{equation}
 \hat{T}_{lij}=\hat{T}_{lij}(z,{\bf k}_\perp)=-ik_lk_ik_j\hat{\psi}. \label{37}
 \end{equation}
With this the $xy$-integration of Eq.~(\ref{33}) becomes
\begin{equation}
G_\perp=\int G_{11}dxdy=\frac{1}{(2\pi)^2}\int \hat{G}_{11}d{\bf k}_\perp,  \label{38}
\end{equation}
where the expression (\ref{25}) for $G_{lq}$ transforms into
\begin{equation}
\hat{G}_{lq}=\hat{T}_{lij}(z, {\bf k}_\perp)\hat{T}_{qij}(z, -{\bf k}_\perp)=k_lk_qk_ik_ik_jk_j\hat{\psi}^2. \label{39}
\end{equation}
Here some care must be taken in the summations as $ik_z$ follows the sign of $z$:
\begin{equation}
 -ik_j\cdot ik_j= k_x^2+k_y^2+(\pm q)^2=k_\perp^2+q^2=2q^2.  \nonumber
 \end{equation}
 With this we find
 \begin{equation}
 \hat{G}_{11}=k_x^2(2q^2)^2\hat{\psi}^2. \label{40}
 \end{equation}
 Symmetry with respect to $x$ and $y$ means that $k_x^2$ can be replaced by $\frac{1}{2}(k_x^2+k_y^2)=\frac{1}{2}k_\perp^2=\frac{1}{2}q^2$ in the integral (\ref{39}), so we get $(z>0)$
 \begin{equation}
 G_\perp =\frac{1}{(2\pi)^2}\int_0^\infty 2q^6\left(\frac{2\pi e^{-qz}}{q}\right)^22\pi qdq=4\pi \frac{5!}{2^6z^6}=\frac{15\pi}{2z^6}. \label{41}
 \end{equation}
By further insertion into Eqs.~(\ref{33}) and (\ref{34}) the results of those integrals are recovered.


\section{General Density}

\subsection{Half-Planes Considered as Composite Particles}

For higher densities, separate oscillators both within each plane and between planes will be correlated. This will add to the complexity of the problem. However, the structure of the problem can be simplified by regarding the two half-planes as a generalized version of a pair of particles. Some details of this approach are given a closer treatment in Appendix A.

The expression (\ref{11}) is a thermal average of four oscillating dipole moments. They have Gaussian distributions since they represent coupled harmonic oscillators. This means that averages can be divided into averages of pairs of dipole moments. Thus we have
\[ g_{ijnm}(\lambda)=\langle s_{1n}(t)s_{2m}(t)s_{1i}s_{2j}\rangle=\langle s_{1n}(t)s_{2m}(t)\rangle \langle s_{1i}s_{2j}\rangle \]
\begin{equation}
+\langle s_{1n}(t)s_{1i}\rangle \langle s_{2m}(t)s_{2j}\rangle +\langle s_{1n}(t)s_{2j}\rangle \langle s_{2m}(t)s_{1i}\rangle. \label{42}
\end{equation}
Now the first term on the right hand side is equal-time average of the operators $A$ and $B$ $\sim s_{1i}s_{2j}$ and should be subtracted from (\ref{9}) to obtain the proper response function. Thus we need
\begin{equation}
\Delta g_{ijnm}(\lambda)=\langle s_{1n}(t)s_{1i}\rangle\langle s_{2m}(t)s_{2j}\rangle+\langle s_{1n}(t)s_{2j}\rangle \langle s_{2m}(t)s_{1i}\rangle. \label{43}
\end{equation}
here the first average represents correlations within the same half-plane while the second is the same for different planes.

To better see the structure or formal contributions to these correlations one may consider the free energy of a pair of one-dimensional oscillators.The harmonic fluctuations result in a distribution function for the dipole moment of each molecule, assuming a static polarizability $\alpha$,
\[ \rho(s)=\exp \left( -\frac{\beta s^2}{2\alpha}\right). \]
If the configuration becomes perturbed by an interaction energy $\phi s_1 s_2$ (not necessarily equal to the Coulomb potential called $\psi$ above), the partition function is
\begin{equation}
Z=\int ds_1ds_2\rho(s_1)\rho(s_2)e^{-\beta \phi s_1 s_2}. \label{44}
\end{equation}
Thus the change in the free energy $F$ due to the mutual interaction $\phi s_1s_2$ becomes\footnote{In Ref.~\cite{brevik88} this expression was used as a basis with $\alpha_2=\alpha_1.$}
\begin{equation}
\ln Z=-\beta F=-\frac{1}{2}\ln (1-\alpha_1\alpha_2\phi^2), \label{45}
\end{equation}where $\alpha_1$ and $\alpha_2$ are the polarizabilities of the two oscillators. The correlation functions $\langle s_a^2\rangle$ ($a=1,2$) and $\langle s_1s_2\rangle$ are
\[\beta\langle s_a^2\rangle=\frac{\alpha_a}{1-\alpha_1 \alpha_2 \phi^2}\]
\begin{equation}
\beta\langle s_1s_2\rangle=\frac{Z'}{Z}=I'=\frac{\alpha_1\alpha_2 \phi}{1-\alpha_1 \alpha_2 \phi^2}, \label{46}
\end{equation}
where the prime means differentiation with respect to $\phi$. Further,
\[ \beta^2(\langle s_1s_2s_1s_2\rangle-\langle s_1s_2\rangle\langle s_1s_2\rangle) =\beta^2(\langle s_1^2\rangle\langle s_2^2\rangle +
\langle s_1 s_2\rangle\langle s_1s_2\rangle )\]
\[ =\frac{Z''}{Z}-\left(\frac{Z'}{Z}\right)^2=I''=\frac{\alpha_1\alpha_2}{1-\alpha_1\alpha_2\phi^2}+\frac{2(\alpha_1\alpha_2\phi)^2}{(1-\alpha_1\alpha_2\phi^2)^2 } \]
\begin{equation}
=\frac{\alpha_1}{1-\alpha_1\alpha_2\phi^2}\,\frac{\alpha_2}{1-\alpha_1\alpha_2\phi^2}+\frac{\alpha_1\alpha_2\phi}{1-\alpha_1\alpha_2\phi^2}\,
\frac{\alpha_1\alpha_2\phi}{1-\alpha_1\alpha_2\phi^2}. \label{47}
\end{equation}
When extended to half-planes one may interpret this expression in the following way: In the first term on the right hand side $\alpha_1$ and $\alpha_2$ are the correlations within each half-plane for $\phi=0$. When the planes interact, the denominator represents induced correlations due to the presence of the second plane. Likewise, the second term in Eq.~(\ref{47}) represents correlations between the planes.

From a fundamental point of view the following feature should however here be noted. The interaction $\phi$ in Eq.~(\ref{47}) will shift the eigenfrequencies of two oscillators further apart. Each term in Eq.~(\ref{47}) will contain both these new frequencies when expanded. For each frequency we can in principle repeat the evaluation that led to Eq.~(\ref{24}). But now $\omega_1 \neq \omega_2$, and the delta function will not contribute. For the situation with low constant relative velocity, there will accordingly be no friction. This is also reasonable since excitation of the quantized system requires disturbances with frequencies that match the energy difference $\hbar (\omega_1-\omega_2)$ while low constant velocity represents the limit of zero frequencies. However, with continuous frequency bands in each oscillator there may again be equal frequencies $\omega_1=\omega_2$ for the perturbed system.

 It is possible to interpret expressions (\ref{45}) - (\ref{47}) in terms of graphs. When expanded, each term in (\ref{45}) represents a closed ring with $\alpha_1$ and $\alpha_2$ vertices that are connected by $\phi$-bonds.  In expression (\ref{46}) for $\langle s_1 s_2 \rangle$ one $\phi$-bond is taken out from the ring
by which one gets a chain of bonds ending on different vertices.  In (\ref{47}) another $\phi$-bond is taken out from the ring which is then split into two chains for which there will be two different situations consistent with the expression on the right hand side of this equation: In one case each chain has both its endpoints on the same particle, while in the other case the chains have their endpoints on different particles.


For our problem we also need correlations in imaginary time, $\lambda=it/\hbar$ as in (\ref{42}) and (\ref{43}). This generalizes Eq.~(\ref{47}) into ($s_a=s_a(0)$)
\[
h(\lambda)=\langle s_1(t)s_2(t)s_1s_2\rangle-\langle s_1(t)s_2(t)\rangle\langle s_1s_2\rangle  \]
\begin{equation}
=h_{11}(\lambda)h_{22}(\lambda)+h_{12}(\lambda)h_{21}(\lambda), \label{48}
\end{equation}
where ($s_{ab}=s_{ba}$)
\begin{equation}
h_{ab}(\lambda)=\langle s_a(t)s_b\rangle. \label{49}
\end{equation}
Its Fourier transform in imaginary time is, similarly to Eq.~(\ref{21}),
\begin{equation}
\tilde{h}(K)=\frac{1}{\beta}\sum_{K_0}\left[ \tilde{h}_{11}(K_0)\tilde{h}_{22}(K-K_0)+\tilde{h}_{12}(K_0)\tilde{h}_{21}(K-K_0)\right]. \label{50}
\end{equation}
In view of the graph interpretation the structure of Eq.~(\ref{50}) will be similar to the one of Eq.~(\ref{47}), so
\begin{equation}
 \tilde{h}_{aa}(K)=\frac{\alpha_{aK}}{1-\alpha_{1K}\alpha_{2K}\phi^2}, \quad (a=1,2) \nonumber
 \end{equation}
\begin{equation}
\tilde{h}_{12}(K)=\frac{\alpha_{1K}\alpha_{2K}\phi}{1-\alpha_{1K}\alpha_{2K}\phi^2}. \label{51}
\end{equation}
With static interactions the $\phi$ will not vary with $K$.

With three-dimensional polarizations the formalism will be modified, but will still be manageable. Some of the points will be verified more explicitly in  Appendix A.

As established in Ref.~\cite{hoye98} the correlation functions for polarizations will be solutions of Maxwell's equations. For dielectric half-planes with permittivities $\varepsilon_1=\varepsilon_1(K)$ and $\varepsilon_2=\varepsilon_2(K)$ one finds (cf. Appendix A)
\begin{equation}
\alpha_{1K}\alpha_{2K}\phi^2 \rightarrow A_{1K} A_{2K} e^{-2qd}, \quad A_{aK}=\frac{\varepsilon_a-1}{\varepsilon_a+1}, \label{52}
\end{equation}
where $d$ is the separation between the planes.

In Eq.~(\ref{51}),  $\alpha_{aK}$ alone represents correlations within each plane. For dilute media it corresponds to the  polarizability whose relation to the permittivity is
\begin{equation}
\varepsilon_a-1 =4\pi \rho_a \alpha_{aK}, \label{53}
\end{equation}
consistent with Eq.~(\ref{52}). As verified in Appendix A it follows that for a general permittivity the $\alpha_{aK}$ in Eq.~(\ref{51}) will generalize to
\begin{equation}
4\pi \rho_a\alpha_{aK} \rightarrow \frac{2(\varepsilon_a-1)}{\varepsilon_a+1}=2A_{aK}, \label{54}
\end{equation}
and by that
\begin{equation}
\phi \rightarrow 2\pi (\rho_1\rho_2)^{1/2}\,e^{-qd}. \nonumber
\end{equation}
Consistent with Eq.~(\ref{52}), the Eq.~(\ref{51}) is modified to
\begin{equation}
2\pi \rho_a\tilde{h}_{aa}(K)=\frac{A_{aK}}{1-A_{1K}A_{2K}e^{-2qd}}, \nonumber
\end{equation}
\begin{equation}
2\pi (\rho_1\rho_2)^{1/2}\tilde{h}_{12}(K)=\frac{A_{1K}A_{2K}e^{-qd}}{1-A_{1K}A_{2K}e^{-2qd}}. \label{55}
\end{equation}
The function $\tilde{h}(K)$ given by Eq.~(\ref{50}) will in the general case replace the $\tilde{g}(K)$ given by Eq.~(\ref{21}). Again one may decompose into contributions from harmonic oscillators as done in Eqs.~(\ref{27})-(\ref{31}). Then Eq.~(\ref{31}) transforms into
\begin{eqnarray}
&&H_0 \rightarrow H_0(u)=\frac{\pi \beta\hbar^2}{2}\int \frac{m^4}{\sinh^2(\frac{1}{2}\beta m)}\nonumber\\
&&\times\Big[ \alpha_{I11}(m^2, u)\alpha_{I22}(m^2,u)
+\alpha_{I12}(m^2,u)\alpha_{I12}(m^2,u)\Big] d\omega, \label{56}
\end{eqnarray}
where
\begin{equation}
\alpha_{Iab}(m^2,u)m^2=-\frac{1}{\pi}\Im (f(-m^2+i\gamma)], \label{57}
\end{equation}
\[ f(K^2)=\tilde{h}_{ab}(K), \quad (a,b=1,2) \]
\[ u=qd. \]
For one particle outside a half-plane one has $\rho_1 \rightarrow 0$ and thus $A_{1K}\rightarrow 0$ (when $\rho_2$ is the density in the half-plane). With this the numerators in the expressions are replaced by 1 and $\tilde{h}_{12}(K)$ vanishes, so in this case the $H_0(u)$ will no longer vary with $x$. The same situation occurs when medium 1 is dilute. Due to this the expressions (\ref{32}) for the friction force will be the same except that in the expression (\ref{31}) for $H_0$ the role of $\alpha_{2K}$ is replaced by $A_{2K}$ as given by Eq.~(\ref{54}). Thus
\begin{equation}
\alpha_{2K}\rightarrow \frac{\varepsilon_2-1}{2\pi \rho_2(\varepsilon_2+1)}, \label{58}
\end{equation}
while $\alpha_{1K}$ is kept.

For larger $\alpha_{1K}$ the $u$-dependence will be present. This will modify the integral (\ref{41}) as $H_0(u)$ has to be included. However, the $z$- and $z_0$- integration over the half-planes that led to the results (\ref{33}) and (\ref{34}) can be evaluated first. With $G_\perp$ given by integral (\ref{41}) we then find ($u=qd$)
\begin{equation}
\int_d^\infty \left( \int_{z>z_0} G_{\perp} dz\right) dz_0 =\frac{\pi}{d^4}\int_0^\infty u^3 e^{-2u}du=\frac{3\pi}{8d^4}. \label{60}
\end{equation}
With this the $H_0$ can be included along with the integrand of (\ref{60}), and the friction force $F$ between two half-planes can again be written in the form (\ref{32}) as
\begin{equation}
F=-GvH_0, \label{61}
\end{equation}
where $G$ again is given by Eq.~(\ref{34}) as $G=3\pi \rho_1\rho_2/(8d^4)$, but where now
\begin{equation}
H_0=\frac{8}{3}\int_0^\infty u^3H_0(u)e^{-2u}du, \label{62}
\end{equation}
with $H_0(u)$ given by Eq.~(\ref{56}).

\subsection{Further Comments on the Complexities Coming from Internal Interactions in the Planes}

Central equations in our context are (\ref{47}) and (\ref{55}). However, their direct use will present certain problems as
the interaction $\phi$ in (\ref{47}) will shift the frequencies of the two oscillators to new eigenfrequencies further apart from each other. Thus when repeating the evaluations that led to Eq.~(\ref{24}) there will no longer be contribution to the friction force at all, since the $\delta$-function will vanish. (Each of the two factors in the two terms of Eq.~(\ref{47}) will then also contain equal frequencies, but they will not contribute to the limiting procedure that led to Eq.~(\ref{24}) \cite{hoye12A}.) So for the situation where the   constant relative velocity is low, there will be no friction. (Continuous frequency bands in each oscillator may change this, but we will not investigate this further here.) This absence of friction may somehow be  reasonable for the following physical reason:  Excitations of the quantized system require disturbances with frequencies that match the energy difference $\hbar(\omega_1 - \omega_2)$, while low constant velocity represents the limit of {\it zero} frequency. Moreover, the expansion (\ref{3}) requires the displacement to be small and comparable to molecular diameters, a situation that may be of minor interest.

Due to these problems we find it more appropriate and realistic to consider the whole mutual interaction $\phi$ as belonging to the perturbing force. An obvious justification for this is that the mutual interaction $\phi=0$ when the two particles are far apart for $t\rightarrow\pm\infty$. Then we are back to our previous situation where the left hand side of Eq.~(\ref{47}) reduces to $\alpha_1 \alpha_2$, i.e.~the nominator in the first term on the right hand side. In the presence of  half-spaces the situation is less trivial, however,  due to interactions within each half-plane. Then the $A_{aK}$ given by Eq.~(\ref{52}) will replace $\alpha_{aK}$ ($a=1,2$). The precise replacement is given by Eq.~(\ref{58}).

 Next, it is noteworthy  that the expression (\ref{58})   can be given a direct physical interpretation. This is most obvious for metals where the dielectric constant is given by the plasma relation ($\varepsilon=\varepsilon_a$)
\begin{equation}
\varepsilon=1-\left(\frac{\omega_p}{\omega}\right)^2,
\label{63}
\end{equation}
the $\omega_p$ being the plasma frequency. Then with Eq.~(\ref{58})
\begin{equation}
2\pi\rho\alpha_K\rightarrow\frac{\varepsilon-1}{\varepsilon+1}
=\frac{\omega_p^2}{\omega_p^2-2\omega^2}
\label{64}
\end{equation}
which corresponds to the polarizability of a harmonic oscillator with eigenfrequency
\begin{equation}
\omega=\frac{\omega_p}{\sqrt{2}}.
\label{65}
\end{equation}
Thus each half-plane represents a set of harmonic oscillators for each wave{\-}vector
${\bf k}_\perp$ in the $xy$-plane. In the idealized situation considered here they all have the same eigenfrequency.

As mentioned above result (\ref{65}) can be given a simple physical interpretation as it represents the frequency of surface plasma waves \cite{barton10B}. Thus each half-plane can be regarded as an assembly of independent harmonic oscillators. There is an oscillator for each wavevector ${\bf k}_\perp$, and they all oscillate with the frequency of surface plasma waves. With these waves there is a net charge that oscillates on the surface while the inside of the halfplane  is neutral. The corresponding electric potential (apart from its time dependence) is given by Eq.~(\ref{A1}) where the coefficients $B$ and $C$ (for $d\rightarrow\infty$) diverges for the frequency of surface plasma waves.

\section{Numerical Examples and Comparison with Earlier Works}


Casimir friction is a complicated topic; it is  approached from different points of view by various researchers, and the obtained results are not always so easy to compare with each other.

The basic method used by us is to employ statistical mechanical methods based upon the Kubo formula. The simplest systems to compare, are clearly those of low particle density. The approach most closely related to ours for such a case, appears to be that of Barton \cite{barton10A,barton10B,barton11}. Even then it turns out that the methods are not easily comparable; cf. the remarks of Barton himself given in  Ref.~\cite{barton11}. In the paper \cite{hoye11B} we showed, however,  by a detailed calculation,  that the results of Barton and those of ours are actually in agreement. This circumstance lends credence to our approach and indicates that we are  on the right track.

The essential new development of the present paper is to extend the statistical mechanical method to the case of finite density. As far as we know, such a treatment has not been given before. Further it is of obvious interest to try to compare our results with  those obtained by others; results that have been derived via  different methods. Results obtained by others are usually based upon  methods of quantum electrodynamics of continuous media, and contain macroscopic concepts such as electrical conductivity.

Now let us consider metal plates for which the permittivity $\varepsilon$ is given by the plasma relation (\ref{63}) as an idealized case. As a physical quantity, the $\varepsilon$ with dispersion will have to include also an imaginary part for $\omega$ real. The simplest generalized version of the expression (\ref{63}) is the one of the Drude model
\begin{equation}
\varepsilon=1+\frac{\omega_p^2}{\zeta(\zeta+\nu)}, \label{100}
\end{equation}
where $\zeta=i\omega$, and where $\nu$ represents damping of plasma oscillations due to finite conductivity of the medium. [Note that a different convention implying $\zeta=-i\omega$ is frequently used, the sign being dictated by the Fourier transform used. We here assume $\tilde{f}(\omega)=\int f(t)e^{-i\omega t} dt$, such that singularities will only be present for $\Im (\omega)>0$, with $f(t)=0$ for $t<0$ due to causality.]

In terms of the variable $K=i\hbar \omega=\hbar \zeta$, Eq.~(\ref{100}) can be written as
\begin{equation}
\varepsilon=1+\frac{2q^2}{K^2+\sigma |K|}, \label{101}
\end{equation}
where
\[ q^2=\frac{(\hbar \omega_p)^2}{2} \quad \rm{and} ~\sigma=\hbar \nu \]
have been introduced. The $\varepsilon$ as function of either $\omega$ or $K$ is the same in the common region where $\varepsilon$ has no singularities, i.e. for $\Im (\omega)<0$ or $\Re(K)>0$ \cite{brevik88}. Further, $\varepsilon(K)$ is symmetric in $K$, and $|K|$ is to be interpreted as $|K|=\lim (K^2+\gamma^2)^{1/2}, \gamma \rightarrow 0$. The singularities representing the singularities of $\varepsilon (K)$ will be along the imaginary $K$ axis or for $K^2$ negative.

For finite density, expression (\ref{58}) is to be formed. So with Eq.~(\ref{101}) for $\varepsilon$, expression (\ref{64}) is modified into
\begin{equation}
2\pi \rho \alpha_K \rightarrow \frac{\varepsilon-1}{\varepsilon+1}=\frac{q^2}{K^2+q^2+\sigma |K|}. \label{102}
\end{equation}
The evaluation of the friction now follows from Eqs.~(\ref{27})-(\ref{34}). By inserting expression (\ref{102}) in Eq.~(\ref{29}), the frequency spectrum is obtained. With $K=im$ we have $(\gamma \rightarrow 0+)$
\begin{equation}
2\pi \rho \alpha_K \rightarrow \frac{q^2(-m^2+q^2-i\sigma m)}{(-m^2+q^2)^2+(\sigma m)^2}, \label{103}
\end{equation}
which gives the frequency spectrum
\begin{equation}
2\pi \rho \alpha_I(m^2)m^2=-\frac{1}{\pi}\Im(f(-m^2))=\frac{q^2}{\pi}\frac{\sigma m}{(-m^2+q^2)^2+(\sigma m)^2}. \label{104}
\end{equation}
For small $\sigma$ we can simplify this, as the expression will be sharply peaked around $m=q$. Thus one might assume to get the main contribution from around this value. However, unless $\beta m$ is small the $\sinh(\cdot)$ term in Eq.~(\ref{31}) will very large, by which values around $m=q$ can be fully neglected; and for small $m$ Eq.~(\ref{104}) can be replaced by
\begin{equation}
m^2\alpha_I(m^2)=Dm, \quad D=\frac{\sigma}{2\pi^2\rho q^2}.
 \label{105}
\end{equation}
This is to be inserted in Eq.~(\ref{31}) to obtain ($\alpha_{I1}=\alpha_{I2}=\alpha_{I}$, $dm=\hbar\,d\omega$)
\begin{eqnarray}
\nonumber
H_0&=&\frac{\pi\beta\hbar}{2}D^2\int\limits_0^\infty\frac{m^2\,dm}{\sinh^2{\frac{1}{2}\beta m}}=\frac{2\pi\hbar}{\beta^2}D^2 I,\\
I&=&\int\limits_0^\infty\frac{x^2 e^{-x}\,dx}{(1-e^{-x})^2}=\sum_{n=1}^\infty\int\limits_0^\infty x^2ne^{-nx}\,dx=2!\sum_{n=1}^\infty\frac{1}{n^2}=\frac{\pi^2}{3}
 \label{106}
\end{eqnarray}
where the substitution $x=\beta m$ has been made.

Finally with Eqs.~(\ref{32}) and (\ref{34}) we find for the friction force per unit area ($\rho_1=\rho_2=\rho,$ and $k_B$ is Boltzmann's constant)
\begin{eqnarray}
\nonumber
F&=&-GvH_0=-\frac{3\pi}{8d^4}\rho^2 vH_0=-\frac{3\pi}{8d^4}\rho^2 v\frac{2\pi\hbar}{\beta^2}D^2I\\
&=&-\frac{v(k_BT)^2\hbar\sigma^2}{16d^4 q^4}=-\frac{\hbar v}{4d^4}\frac{(k_B T)^2(\hbar\nu)^2}{(\hbar\omega_p)^4}.
 \label{108}
\end{eqnarray}

 Let us consider a numerical example.  Assume room temperature, $T=300$ K, corresponding to $k_BT=25.86$ meV. Choose gold as medium, for which $\hbar \omega_p=9.0$\,eV and $\hbar \nu=35$\,meV. Then  choose $v=100$ m/s for the relative velocity and a small separation $d=10$ nm between the plates. With $\hbar=1.054\cdot10^{-34}$ Js we then find for the friction force (\ref{108})
\begin{equation}
F= 3.29\cdot10^{-11}\,\mbox{Pa}.
\label{109}
\end{equation}
This is a very small force. However, by changing parameters this will change rapidly by which $F$ can become very large instead. But first let us compare this force with the result obtained by Pendry \cite{pendry97} for $T=0$ where the friction linear in velocity was found to be zero, assuming constant conductivity.  Instead a non-zero force, proportional to $v^3$,  was  found. The influence of relative velocity is to create an oscillating force between the particles; these   oscillations will create excitations from the $T=0$ ground states of low frequency oscillations and thus contribute to friction. According to the derivations of Ref.~\cite{pendry97} the friction for low and not too high velocities for a dielectric function $\varepsilon=1+i\sigma/{\omega\varepsilon_0}$ was found to be
\begin{equation}
F_P=\frac{5\hbar\varepsilon_0^2 v^3}{2^8\pi^2\sigma^2d^6}
\label{110}
\end{equation}
(here and henceforth $\sigma$ is the conductivity in SI units, not $\hbar \nu$ as above).
When compared with our dielectric function Eq.~(\ref{100}) for small $\omega$ one sees that the $\sigma/\varepsilon_0$ of this equation is our $\omega_p^2/\nu$. This can then be inserted in Eq.~(\ref{110}), and one finds the ratio between the friction forces (\ref{108}) and (\ref{110}) to be
\begin{equation}
\frac{F}{F_P}=\frac{64\pi^2}{5}\left(\frac{k_B T}{\hbar v/d}\right)^2.
\label{111}
\end{equation}
This simple expression can be given a direct physical interpretation. The $k_B T$ is the energy quanta that can be excited due to thermal energies or fluctuations while $\hbar v/d$ are the energy quanta generated due to frequencies $v/d$ generated by the finite velocity. Due to the physical interpretation above and arguments used in Ref.~\cite{pendry97}, there is reason to expect its result for finite velocity and zero temperature to be consistent with ours, obtained for small velocity and nonzero temperature.
For the numerical values that gave the force (\ref{109}) one finds the ratio
\begin{equation}
\frac{F}{F_P}=1.95\cdot10^9.
\label{112}
\end{equation}
Thus thermal energies have a much greater influence that the frequencies generated from the finite velocity. In Ref.~\cite{pendry97}, a much greater friction was obtained by using other numerical values. There $\sigma=0.1$\,$\Omega^{-1}$m$^{-1}$, $d=10^{-10}$\,m, and $v=1.0$\,m/s was used. With permittivity of vacuum $\varepsilon_0=8.85\cdot10^{-12}$\,As/Vm one then finds $\omega_p^2/\nu=\sigma/\varepsilon_0=1.12\cdot10^{10}$\,s$^{-1}$ while result (\ref{109}) was based upon  $\sigma/\varepsilon_0=\omega_p^2/\nu=3.5\cdot10^{18}$\,s$^{-1}$, a much greater number. The numbers of Ref.~\cite{pendry97} will thus give the very large friction force
\begin{equation}
F=3.5\cdot10^{12}\,\rm{Pa}
\label{112}
\end{equation}
which with the ratio (\ref{111}) means (as $v/d$ is unchanged) $F_P=1.6\cdot10^3$\,Pa. In Ref.~\cite{pendry97} the result $F_P\approx 3\cdot 10^3$\,Pa was found by substitution into its high velocity formula (the formula least sensitive to velocity). Anyway, the latter force (\ref{112}) is unrealistically large. One reason is that a separation $d=10^{-10}$\,m would more or less imply direct contact between the particles of the two half-planes. Further the force decreases very rapidly with increasing separation.

Another factor of influence here  is that the effective separation between the planes will increase when the charge density of electrons becomes small. This has not been taken into account. Clearly, when neglecting dielectric effects otherwise, the force should vanish when the density of electrons and thus $\sigma$ vanishes while results (\ref{108}) and (\ref{111}) tell the opposite. In this respect the Casimir force between parallel plates filled with a plasma of electrons was considered by
us earlier (in the classic electrostatic limit) \cite{hoye08,hoye09}. Then it was found that the separation $d$ increased to an effective separation $d+2/\kappa$ (for small densities) where $\kappa$ is the inverse shielding length where $\kappa^2=4\pi\beta e^2 \rho$ ($e$ is charge of electron) \cite{hoye08}.


Among  other previous approaches with which it seems natural to compare our results, we shall focus on those  of Volokitin and Persson. As mentioned earlier, they have written a series of papers on this topic \cite{volokitin99,volokitin03,volokitin07,volokitin08,volokitin11}. In contrast  to the paper of Pendry \cite{pendry97}, they considered finite temperatures.
In their review article \cite{volokitin07} a variety of situations were considered. One such situation is for parallel relative motion of metal plates. The friction coefficient according to their Eq.~(97) is then
\begin{equation}
\gamma_{\| p}^{\rm evan}\approx0.3\frac{\hbar}{d^4}\left(\frac{k_B T}{4\pi\hbar\sigma}\right)^2
\label{113}
\end{equation}
where here the conductivity $\sigma$ is given in Gaussian units such that $4\pi\sigma=\omega_p^2/\nu$. This follows from their Eq.~(38) which is proportional to our expression (\ref{104}). The friction force $F_{VP}=\gamma_{\| p}^{\rm evan}v$ is thus directly comparable to our result (\ref{108}), and we find the ratio
\begin{equation}
\frac{F_{VP}}{F}\approx 1.2.
\label{114}
\end{equation}
Thus we can conclude that the two expressions for the friction force are consistent except for a small difference in numerical prefactor.

 A closer look at integral (92) for $\gamma_{\| p}^{\rm evan}$ in Ref.~\cite{volokitin07} makes it  probable that this difference is due to the term $1/(1-u)^2=\sum_{n=1}^\infty nu^{n-1}$ in the integrand with $u=e^{-2|\gamma|d}$. This term is for small $\omega\rightarrow 0$ for which the reflection coefficients $R_{ip}\rightarrow 1$ ($i=1,2$), i.e.~$(\varepsilon-1)/(\varepsilon+1)\rightarrow 1$. (Actually, when  comparing with Eq.~(87) of that reference and with Eq.~(27) of Ref.~\cite{pendry97} the $|\gamma|$ seems to be a misprint for the integration variable $q$.) Then one has integrals of the form $\int_0^\infty nu^nq^3\,dq\propto n^{-3} d^{-4}$. This gives the sum
\begin{equation}
\sum_{n=1}^\infty\frac{1}{n^3}=\zeta(3)=1.202\cdots
\label{115}
\end{equation}
where $\zeta(z)$ is the $\zeta$-function. This sum may be precisely the ratio (\ref{114}). Here it can be noted that the term above is also present in our expressions (\ref{47}) and (\ref{55}) with $u=\alpha_1 \alpha_2 \phi^2$ and $u=A_{1K}A_{2K} e^{-2qd}$ respectively, but was later disregarded by further explicit evaluations as discussed in Sec.~4.2.

In view of this, observing the similarity of the expressions, the results of Ref.~\cite{volokitin07} seem consistent with our results as they agree numerically except from some uncertainty in the  prefactor. 

\vspace{2cm}

\renewcommand{\theequation}{\mbox{\Alph{section}.\arabic{equation}}}
\appendix

\setcounter{equation}{0}
\section{Background for the expressions for the correlation functions}

Here we will justify expressions (\ref{55}) for the correlation functions and the arguments that led to them. As discussed  in Ref.~\cite{hoye98}, the correlation functions can be identified as solutions of Maxwell's equations. In \cite{hoye98} the solution was derived for two half-planes with equal permittivities. Here we will consider two half-planes with permittivities $\varepsilon_1$ and $\varepsilon_2$, separated by a distance $d$. For the electrostatic case, which we will assume, the Coulomb interaction for a point charge as given in the form (\ref{36}) will be the basis.

Let the half-planes be parallel to the $xy$-plane with surfaces at $z=0$ and $z=d~(>0)$. Their permittivities are $\varepsilon_1$ for $z<0$ and $\varepsilon_2$ for $z>d$. With a unit charge located at $z=z_0<0$ the resulting potential can be written as $(q=k_z)$

\begin{equation}
\hat{\psi}(z, k_\perp)=\frac{2\pi}{k_\perp}e^{qz_0}\left\{ \begin{array}{llll}
\frac{1}{\varepsilon_1}e^{-2qz_0}e^{qz}+Be^{qz}, & z<z_0 \\
\frac{1}{\varepsilon_1}e^{-qz}+Be^{qz}, & z_0<z<0 \\
Ce^{-qz}+C_1e^{qz}, & 0<z<d \\
De^{-qz}, & d<z.
\end{array}
\right. \label{A1}
\end{equation}
At the boundaries the $\hat \psi$ and its derivatives $\varepsilon_a\partial \hat{\psi}/\partial z$ $(a=1,2)$ should be continuous. This yields the equations
\begin{equation}
\frac{1}{\varepsilon_1}+B=C+C_1, \nonumber
\end{equation}
\begin{equation}
\varepsilon_1 \left( \frac{1}{\varepsilon_1}-B\right) =C-C_1, \nonumber
\end{equation}
\begin{equation}
Ce^{-qd}+C_1e^{qd}=De^{-qd}, \nonumber
\end{equation}
\begin{equation}
Ce^{-qd}-C_1e^{qd}=\varepsilon_2De^{-qd}. \label{A2}
\end{equation}
These equations may first be solved for $C$ and $C_1$ in terms of $D$. Then solving for $D$ and $B$ one finds
\begin{equation}
D=\frac{4}{(\varepsilon_1+1)(\varepsilon_2+1)\left( 1-A_{1K}A_{2K}e^{-2qd}\right)}, \nonumber
\end{equation}
\begin{equation}
B=\frac{1}{\varepsilon_1}A_{1K}-\frac{\varepsilon_2-1}{\varepsilon_1+1}De^{-2qd} \label{A3}
\end{equation}
with $(a=1,2)$
\begin{equation}
A_{aK}=(\varepsilon_a-1)/(\varepsilon_a+1).
\label{A4}
\end{equation}

One notes that the denominator of expression (\ref{55}) is the one of $D$. Further Eq.~(\ref{A4}) verifies expression (\ref{54}). However, to fully verify the expressions given by Eq.~(\ref{55}) and thus (\ref{56}) the $\hat\psi$ of Eq.~(\ref{36}) should be replaced with the $\hat\psi$ of Eq.~(\ref{A1}). To do this in detail the general form of the pair correlation function for a uniform dielectric fluid as found by H{\o}ye and Stell can be used \cite{hoye76}. As pointed to in Eq.~(5.5) in Ref.~\cite{hoye76} (cf. also the more detailed considerations in Ref.~\cite{hoye98}), this includes a prefactor $((\varepsilon -1)/(3y))^2$ with $3y=4\pi\rho\alpha_K$. With endpoints in separate half-planes this should generalize  to $((\varepsilon -1)/(3y))^2 \rightarrow((\varepsilon_1-1)/(3y_1))((\varepsilon_2-1)/(3y_2))$ where $\alpha_K\rightarrow\alpha_{aK}$ $(a=1,2)$. This is then integrated with the $T_{lij}$ of Eq.~(\ref{37}). One will find that such an evaluation is rather nontrivial. Thus we will not try to perform it here. Instead we note that the result of all this should merely produce the factors $A_{1K}$ and $A_{2K}$ as shown in the numerator of Eq.~(\ref{55}). This will follow by the arguments given in the paragraph above Eq.~(\ref{48}) where the expressions for a pair of oscillators are interpreted in terms of graphs with vertices and $\phi$-bonds. In this respect the pair of half-planes can be regarded as a generalization of a pair of point particles.

\setcounter{equation}{0}
\section{Remark on  a formal relationship to  quantum field theory}

As we stated above, the Fourier transform $\tilde{\phi}(\omega)$ of the response function $\phi(t)$ is the same as the Fourier transform $\tilde{g}(K)$ of the correlation function $g(K)$ at imaginary time, $\lambda=it/\hbar$. The equality reads,  Eq.~(\ref{15}) being here reproduced for convenience,
\begin{equation}
\tilde{\phi}(\omega)=\tilde{g}(K), \label{B1}
\end{equation}
where  $K=i\hbar \omega$ is the imaginary frequency.

It turns out that the quantum statistical mechanics for particles, and the quantum theory for fields, are closely related although the correspondence is not always so easy to see from a mere inspection. Therefore, we found it useful to point out how a parallel to Eq.~(\ref{B1}) reads in the conventional QFT for the electromagnetic field.

Consider, for definiteness, Schwinger's source theory in the form presented, for instance, in Ref.~\cite{schwinger78}. For simplicity, as common in field theory, we work with natural units so that $\hbar=c= k_B=1$. The electric field components $E_i(x)$ are related to the polarization components $P_k(x')$ via a tensor $\Gamma_{ik}$, called the generalized susceptibility,
\begin{equation}
E_i(x)=\int d^4x' \Gamma_{ik}(x,x')P_k(x'), \label{B2}
\end{equation}
where $x=({\bf r},t)$. Stationarity of the system means that $\Gamma_{ik}$ depends on time only through the difference $\tau=t-t'$. Causality implies that the integration over $t'$ is limited to $t' \leq t$.

Introduce the Fourier transform $ \Gamma_{ik}({\bf r, r'},\omega)$ via
\begin{equation}
\Gamma_{ik}(x,x')=\int_{-\infty}^\infty \frac{d\omega}{2\pi}e^{-i\omega \tau}\Gamma_{ik}({\bf r,r'},\omega). \label{B3}
\end{equation}
The function $\Gamma_{ik}({\bf r,r'},\omega)$  is known to be one-valued in the upper half frequency plane; it has no singularity on the real axis (omitting metals), and it does not take real values at any finite point in the upper half plane except on the imaginary axis.

From Kubo's formula we can now write
\begin{equation}
\Gamma_{ik}({\bf r,r'},\omega)=i\int_0^\infty d\tau e^{i\omega \tau}\langle [E_i(x), E_k(x')]\rangle. \label{B4}
\end{equation}
It means that the generalized susceptibility can be identified with the retarded Green function: $\Gamma_{ik}(x,x')=G_{ik}^R(x,x')$. For $t<t'$ both $\Gamma_{ik}(x,x')$ and $G_{ik}^R(x,x')$ vanish, the first because of causality, the second because of the definition of the retarded Green function. [Note as commented below Eq.~(\ref{100}) the Fourier transform in this section corresponds to the other convention $\zeta =-i\omega$.]

Consider now the the correlation $\langle E_i(x)E_k(x')\rangle$. Its Fourier transform $\langle E_i({\bf r},\omega)E_k({\bf r'},\omega')\rangle$ (in field theory commonly called the two-point function) can be expressed in terms of the spectral correlation $\langle E_i({\bf r})E_k({\bf r'})\rangle_\omega$ as
\begin{equation}
\langle E_i({\bf r},\omega)E_k({\bf r'},\omega')\rangle = 2\pi \langle E_i({\bf r})E_k({\bf r'})\rangle_\omega \delta(\omega+\omega'). \label{B5}
\end{equation}
Now make use of the fluctuation-dissipation theorem \cite{landau80} to get
\begin{equation}
\Im \,G_{ik}^R({\bf r,r'},\omega)\coth \left( \frac{1}{2}\beta \omega \right)= \langle E_i({\bf r})E_k({\bf r'})\rangle_\omega, \label{B6}
\end{equation}
with $\beta=1/T$. Equation (\ref{B6}) is the field-theoretical counterpart of Eq.~(\ref{B1}). In both cases we see that there is  a close relationship between the response (or Green function) and the correlation.

To make the connection to the statistical mechanical method more explicit, one may consider expressions (\ref{B3}) - \ref{B6}) for an oscillator which has a frequency spectrum. Its response function (\ref{6}) is
\begin{equation}
\phi(t)=\frac{1}{i\hbar}\rm{Tr}\{\rho[s(0), s(t)]\}
\label{B7}
\end{equation}
where $s(t)$ is polarization or amplitude. The $\phi(t)$ corresponds to the $\Gamma_{ik}(x,x')$ of Eq.~(\ref{B3}) while the right hand side of Eq.~(\ref{B7}) corresponds to the integrand of (\ref{B4}). Further the Fourier transform $\tilde\phi(\omega)$ corresponds to $\Gamma_{ik}({\bf r},{\bf r'},\omega)$ with $\omega\rightarrow-\omega$ due to definition (\ref{16}) of $\tilde\phi(\omega)$.

Now for a simple harmonic oscillator expression (\ref{22}) for $\tilde g_a(K)\rightarrow\tilde g(K)\rightarrow\tilde\phi(\omega)$ is valid. The inverse transform (\ref{17}) to imaginary time $\lambda=it/\hbar$ will for this $\tilde g(K)$ give ($\omega_a\rightarrow\omega_0$)
\begin{equation}
g(\lambda)=\langle s(t)s(0) \rangle_{\omega_0} =\frac{1}{2}\alpha\hbar\omega_0\frac{\cosh(\frac{1}{2}\beta\hbar\omega_0-\lambda)}{\sinh{(\frac{1}{2}\beta\hbar\omega_0)}},
\label{B8}
\end{equation}
\begin{equation}
g(0)=\langle s(0)s(0) \rangle_{\omega_0} =\frac{1}{2}\alpha\hbar\omega_0\coth{(\frac{1}{2}\beta\hbar\omega_0)}.
\label{B9}
\end{equation}
With a distribution of eigenfrequencies Eqs.~(\ref{27}) and (\ref{29}) are valid. So from Eqs.~(\ref{28}) and (\ref{29}) one finds ($m=\hbar\omega_0$)
\begin{equation}
g(0)=\frac{1}{2\pi}\int\limits_{-\infty}^\infty\langle s(0)s(0) \rangle_{\omega_0}\,d\omega_0 =\int\limits_0^\infty\alpha_I(m^2)\frac{1}{2}m\coth{(\frac{1}{2}\beta m)}\,d(m^2)
\label{B10}
\end{equation}
where with $K=i\hbar\omega_0$
\begin{equation}
\alpha_I(m^2)m^2=-\frac{1}{\pi}\Im [g(K)]=-\frac{1}{\pi}\Im[\phi(\omega_0)].
\label{B11}
\end{equation}
Now
\begin{equation}
\int\limits_0^\infty m\,d(m^2)=2\int\limits_0^\infty m^2\,dm=\int\limits_{-\infty}^\infty m^2\,dm,
\label{B12}
\end{equation}
so
\begin{equation}
g(0)=-\frac{1}{2\pi}\int\limits_0^\infty\Im [\phi(\omega_0)]\coth{(\frac{1}{2}\beta m)}\,dm.
\label{B13}
\end{equation}
Thus with Eqs.~(\ref{B10}) and (\ref{B13}) one finds ($m=\omega_0$, $\hbar=1$)
\begin{equation}
-\Im[\phi(\omega_0)]\hbar\coth{(\frac{1}{2}\beta m)}=\langle s(0)s(0) \rangle_{\omega_0}
\label{B14}
\end{equation}
which corresponds to Eq.~(\ref{B6}) of field theory. (The minus sign is due to the shift $\omega\rightarrow -\omega$ by Fourier transform.)

However, by the statistical mechanical approach Eq.~(\ref{B6}) is not of primary interest. Instead relation (\ref{B1}), which is independent of temperature for harmonic oscillators, is crucial. Further the electromagnetic field, which is quantized as a set of harmonic oscillators, can be eliminated to be replaced by its Green function $\phi(\omega)=g(K)$ that acts as a pair interaction between induced dipole moments.

\bibliographystyle{mdpi}
\makeatletter
\renewcommand\@biblabel[1]{#1. }
\makeatother

\end{document}